\documentclass[aps,prl,reprint,superscriptaddress]{revtex4-1}

\usepackage{mathrsfs}
\usepackage{amsmath}
\usepackage{graphicx}
\usepackage{multirow}

\begin{document}


\title{Observation of anti-$\mathcal{PT}$ symmetry phase transition in the magnon-cavity-magnon coupled system}


\author{Jie Zhao}
\altaffiliation{These authors contributed equally to this work}
\affiliation{Hefei National Laboratory for Physical Sciences at the Microscale and Department of Modern Physics, University of Science and Technology of China, Hefei 230026, China
}
\affiliation{CAS Key Laboratory of Microscale Magnetic Resonance, University of Science and Technology of China, Hefei 230026, China}
\affiliation{Synergetic Innovation Center of Quantum Information and Quantum Physics, University of Science and Technology of China, Hefei 230026, China}

\author{Yulong Liu}
\altaffiliation{These authors contributed equally to this work}
\affiliation{Department of Applied Physics, Aalto University, P.O. Box 15100, FI-00076 Aalto, Finland}

\author{Longhao Wu}
\affiliation{Hefei National Laboratory for Physical Sciences at the Microscale and Department of Modern Physics, University of Science and Technology of China, Hefei 230026, China
}
\affiliation{CAS Key Laboratory of Microscale Magnetic Resonance, University of Science and Technology of China, Hefei 230026, China}
\affiliation{Synergetic Innovation Center of Quantum Information and Quantum Physics, University of Science and Technology of China, Hefei 230026, China}

\author{Chang-Kui Duan}
\affiliation{Hefei National Laboratory for Physical Sciences at the Microscale and Department of Modern Physics, University of Science and Technology of China, Hefei 230026, China
}
\affiliation{CAS Key Laboratory of Microscale Magnetic Resonance, University of Science and Technology of China, Hefei 230026, China}
\affiliation{Synergetic Innovation Center of Quantum Information and Quantum Physics, University of Science and Technology of China, Hefei 230026, China}

\author{Yu-xi Liu}
\affiliation{Institute of Microelectronics, Tsinghua University, Beijing 100084, China}

\author{Jiangfeng Du}
\email[]{djf@ustc.edu.cn}
\affiliation{Hefei National Laboratory for Physical Sciences at the Microscale and Department of Modern Physics, University of Science and Technology of China, Hefei 230026, China
}
\affiliation{CAS Key Laboratory of Microscale Magnetic Resonance, University of Science and Technology of China, Hefei 230026, China}
\affiliation{Synergetic Innovation Center of Quantum Information and Quantum Physics, University of Science and Technology of China, Hefei 230026, China}


\date{\today}

\begin{abstract}
As the counterpart of $\mathcal{PT}$ symmetry, abundant phenomena and potential applications of anti-$\mathcal{PT}$ symmetry have been predicted or demonstrated theoretically. However, experimental realization of the coupling required in the anti-$\mathcal{PT}$ symmetry is difficult. Here, by coupling two YIG spheres to a microwave cavity, the large cavity dissipation rate makes the magnons coupled dissipatively with each other, thereby obeying a two-dimensional anti-$\mathcal{PT}$ Hamiltonian. In terms of the magnon-readout method, a new method adopted here, we demonstrate the validity of our method in constructing an anti-$\mathcal{PT}$ system and present the counterintuitive level attraction process. Our work provides a new platform to explore the anti-$\mathcal{PT}$ symmetry properties and paves the way to study multi-magnon-cavity-polariton systems.
\end{abstract}

\pacs{xxxxxx}

\maketitle

In the real world, quantum systems interact with surrounding environment and evolve from being closed originally into open ones \cite{de2017dynamics, xiang2013hybrid}. Hamiltonians describing open systems are generally non-Hermitian. Due to the non-conserving nature, the eigen-energies are complex numbers and the corresponding dynamics are complicated \cite{miri2019exceptional}. One special type of non-Hermitian systems, which respects parity-time ($\mathcal{PT}$) symmetry, has triggered unprecedented interest, and been widely explored theoretically and experimentally \cite{el2018non, feng2017non, ozdemir2019parity, Wu878observation}. Another special type of non-Hermitian systems is the anti-$\mathcal{PT}$ symmetric system, which is the counterpart of the $\mathcal{PT}$ symmetric system and always preserve conjugated properties to those observed in $\mathcal{PT}$-symmetric ones \cite{peng2016anti, choi2018observation}. Based on the conjugated properties, abundant phenomena and potential applications of anti-$\mathcal{PT}$ systems have been predicted or demonstrated theoretically. Examples include unidirectional light propagation \cite{lau2018fundamental}, flat full transmission bands \cite{yang2017anti}, enhanced sensor sensitivity \cite{wiersig2014enhancing}, constructing topological superconductor \cite{gong2018topological}, and potential effects on quantum measurement back-action evading \cite{bernier2018level}. Motivated by the intriguing phenomena and various potential applications, experimental realizations of anti-$\mathcal{PT}$ symmetric systems are highly desirable. However, because of the requirement of purely imaginary coupling constants between two bare states, there have been few experimental works about anti-$\mathcal{PT}$ symmetry \cite{peng2016anti, choi2018observation, zhang2018dynamically, li2019anti}.

Recently, collective excitations of spin ensembles in ferromagnetic systems (also called as magnons) have drawn considerable attentions due to their very high spin density, low damping rate, and high-cooperativity with the microwave photons \cite{lachance2019hybrid, goryachev2018cavity}. Especially the ferromagnetic mode in an yttrium iron garnet (YIG) sphere can strongly \cite{goryachev2014high, zhang2014strongly, tabuchi2014hybridizing, zhang2015cavity} and even ultra-strongly \cite{zhang2016superstrong, bourhill2016ultrahigh} couple to the microwave cavity photons, leading to cavity-magnon polaritons. Based on cavity-magnon polaritons, quantum memories have been realized \cite{zhang2015magnon}, remote coherent coupling between two magnons has been proposed \cite{rameshti2018indirect} and observed \cite{lambert2016cavity}. At the same time, coupled cavity-magnon polaritons are attractive systems for exploring non-Hermitian physics \cite{zhang2019higher, zhang2018dynamically, zhang2017observation}, because of their easy reconfiguration, flexible tunability, and especially the strong compatibility with microwave \cite{tabuchi2015coherent, lachance2017resolving}, optics \cite{kusminskiy2016coupled, hisatomi2016bidirectional, zhang2016optomagnonic, graf2018cavity}, as well as mechanical resonators \cite{holanda2018detecting, zhang2016cavity}.

Here, we propose a coupled magnon-cavity-magnon polariton system to experimentally demonstrate the anti-$\mathcal{PT}$ symmetry \cite{yang2017anti}. The pure imaginary coupling between two spatially separated and frequency detuned magnon modes is realized by engineering the dissipative reservoir of the cavity field. Different from previous cavity-magnon-polariton experiments, in which the signals are extracted from cavity \cite{goryachev2014high, zhang2014strongly, tabuchi2014hybridizing, zhang2015cavity, zhang2016superstrong, bourhill2016ultrahigh}, we need to extract the signals from the magnons. The experimental data are not only fitted well with the original experimental Hamiltonian calculated transmission spectrum but also the one predicted by the standard anti-$\mathcal{PT}$ Hamiltonian. By continuously tuning non-Hermitian control parameter, e.g. cavity decay rate, we present the spontaneous symmetry-breaking transition, which is accompanied by the energy level attraction. The results are compared with the data normally obtained from the cavity. This comparison demonstrates that the magnon-readout technique enables us to measure the magnon state separately in a multi-magnon-cavity coupled system and allow the exploration of many significant phenomena.

\begin{figure}
\includegraphics[width=\linewidth]{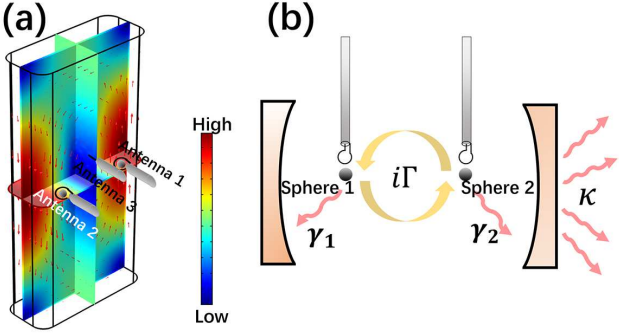}
\caption{\label{fig1}(color online) (a). The schematic diagram of our experimental system. Two YIG spheres are placed inside an oxygen free copper made 3D cavity. Antenna 1 and antenna 2 are coupled to the YIG spheres, and antenna 3 is coupled with the cavity. These three antennae can be connected to a network analyzer (VNA) to measure the transmission spectra $\rm S_{11}$, $\rm S_{22}$ and $\rm S_{33}$. In the experiment, antenna 3 can be used to control the dissipation rate of the cavity. The colored slice figure shows the simulated magnetic field distribution of the cavity $\rm TE_{101}$ mode. (b). The coupling mechanism between two YIG spheres. When the cavity dissipation rate $\kappa$ is much larger than the dissipation rates of the two magnons, i.e., $\kappa \gg \gamma_1,\ \gamma_2$, the two magnons are dissipatively coupled with each other and the cavity behaves as a dissipative coupling medium.}
\end{figure}

Our experimental setup is schematically shown in FIG.~\ref{fig1} (a). Two YIG spheres are placed inside a three-dimensional (3D) oxygen-free copper cavity with inner dimensions $40\times 20\times 8\ \rm mm^3$. The YIG spheres with 0.3 mm diameter are glued on one end of two glass capillaries, which are anchored at two mechanical stages. The YIG spheres are placed near the magnetic-field antinode of the cavity mode $\rm TE_{101}$ through two holes in the cavity wall. Two grounded loop readout antennae, antenna 1 and antenna 2, are coupled with the YIG sphere 1 and sphere 2, respectively. In this setup, we can change the position of YIG spheres relative to loop antennae by tuning the mechanical stages. In our experiment, we focus on the Kittle mode, which is a spatially uniform ferromagnetic mode. To avoid involving other magnetostatic modes, the antennae are carefully designed and assembled. The antenna 3 with a length tunable pin is used to control the dissipation rate of the cavity. When we probe the system from the cavity, the antenna 1 and antenna 2 are removed. The whole system is placed in a static magnetic bias field, which is created by a high-precision room temperature electromagnet. The bias magnetic field and the magnetic field of the $\rm TE_{101}$ cavity mode are nearly perpendicular at the site of two YIG spheres.

In our system, the two YIG spheres work at low excitation regime, thus the collective spin excitation of YIG spheres can be simply regarded as harmonic resonators. In dissipative regime, our system can be approximately described by the standard anti-$\mathcal{PT}$ Hamiltonian \cite{yang2017anti} (Supplementary Materials A):

\begin{equation}\label{eff-Hamiltonian}
H_{\rm eff} = \left[
\begin{matrix}
\Omega-i(\gamma + \Gamma) & -i \Gamma \\
-i\Gamma & -\Omega-i(\gamma + \Gamma)
\end{matrix}
\right].
\end{equation}

\begin{table*}
\setlength{\tabcolsep}{4mm}{
\centering
\caption{\label{sys param} Parameters used in cavity-readout and magnon-readout methods. $\gamma_1$ and $\gamma_2$ are the dissipation rates of magnon 1 and magnon 2, respectively. $g_{13}$ or $g_{23}$ is the coupling strength between the cavity and the magnon 1 or magnon 2. $\left|\Omega\right|$ is the effective detuning. $\kappa_{\rm int}$ is the intrinsic dissipation rate of the cavity (without additional ports). $\kappa_1$, $\kappa_2$ and $\kappa_3$ are the dissipation rates introduced by antenna 1, antenna 2 and antenna 3, respectively. }
\begin{tabular}{|c|c|c|c|c|c|c|c|c|c|}
\hline
\multicolumn{1}{|c|}{ \multirow{2}*{Probe Method} }& \multicolumn{9}{|c|}{System Parameters (units: $2\pi\times\ \rm MHz$)}\\
\cline{2-10}
\multicolumn{1}{|c|}{}&$\gamma_1$&$\gamma_2$&$g_{13}$&$g_{23}$&$\left|\Omega\right|$&$\kappa_{\rm int}$ & $\kappa_1$ & $\kappa_2$ & $\kappa_3$\\
\hline
\multicolumn{1}{|c|}{Cavity-readout}&$1.11$&$1.11$&$9.77$&$9.61$&$2.7$&tunable&0&0&$\approx\kappa_{\rm int}$\\
\hline
\multicolumn{1}{|c|}{Magnon-readout}&$2.22$&$2.22$&$6.65$&$6.41$&$2.7$&$1.5$&$0.45$&$0.92$&tunable\\
\hline
\end{tabular}}
\end{table*}

\noindent Here $i\Gamma$ is the dissipative coupling rate, $\Omega = (\omega_1 - \omega_2)/2$ is the effective detuning in the rotating reference frame with frequency $(\omega_1 + \omega_2)/2$, where $\omega_{1}$ ($\omega_2$) is the resonant frequency of magnon 1 (2). For the Kittle mode, the frequency of a magnon linearly depends on the bias field $\vec{B}_{i}$, i.e., $\omega_{i} = \gamma_0\left|\vec{B}_{i}\right|+\omega_{m, 0}\ (i=1,\ 2)$, where $\gamma_0=28\ \rm GHz/T$ is the gyromagnetic ratio and $\omega_{m, 0}$ is determined by the anisotropy field. To obtain the effective Hamiltonian in Eq. (\ref{eff-Hamiltonian}), we further require that the dissipation rates of two magnons are nearly equal, i.e., $\gamma_1 \approx \gamma_2 = \gamma$, and the magnon 1 - cavity coupling rate $g_{13}$ approximately equals to the magnon 2 - cavity coupling rate $g_{23}$, i.e., $g_{13} \approx g_{23} = g$. In the regime of $\kappa \gg \gamma$ and $\kappa \gg \left| \omega_3 - \omega_{1(2)} \right|$, with the cavity dissipation rate $\kappa$, the effective coupling rate is $\Gamma = g^2/\kappa$. We can conveniently obtain the eigenvalues of the Hamiltonian in Eq.~(\ref{eff-Hamiltonian}), $\lambda_{\pm} = -i(\gamma+\Gamma)\pm \sqrt{\Omega^2-\Gamma^2}$. When $\left|\Omega\right|>\left|\Gamma\right|$, the eigenvalues are normally complex, and the system works in anti-$\mathcal{PT}$ symmetry broken phase regime. If $\left|\Omega\right|<\left|\Gamma\right|$, the eigenvalues are purely imaginary and the system works in anti-$\mathcal{PT}$ symmetry phase regime. The condition of $\left|\Omega\right|=\left|\Gamma\right|$ defines the EP.

We can probe the magnon-cavity-polariton system from either the magnon or the cavity. When we probe the system from the magnon, we carefully tune the mechanical stage and change the position of YIG spheres relative to the readout antennae to change the external dissipation rate $\gamma_{11}$ ($\gamma_{21}$) of magnon 1 (2), so that the readout antennae are critically coupled to the magnons, i.e., $\gamma_{i0} \approx\gamma_{i1}$ ($i=1,\ 2$), where $\gamma_{i0}$ is the intrinsic dissipation rate of magnon. In this situation, the total dissipation rate of magnon 1 (2) should be $\gamma_{i}=\gamma_{i0}+\gamma_{i1}\approx 2\gamma_{i0}$ ($i=1,\ 2$). In this setup, the dissipation rate of the cavity is controlled by solely changing the pin length of the antenna 3. When we probe the system from the cavity, the signal is injected into the cavity from antenna 3, and the reflected signal is measured from the same port. In this case, the overall dissipation rate of magnon 1 (2) equals to the intrinsic dissipation rates, i.e., $\gamma_{i}=\gamma_{i0}$ ($i=1,\ 2$). In this measurement setup, we require that antenna 3 is critically coupled to the cavity. To accomplish this requirement, we paste carbon tape at the electric-field antinode of the cavity mode to change the cavity intrinsic dissipation rate $\kappa_{\rm int}$ and change the pin length of antenna 3 to change the dissipation rate $\kappa_3$, such that the condition $\kappa_{\rm int} \approx\kappa_3$ can be satisfied. All system parameters used in both readout methods are presented in TABLE \ref{sys param}, which shows that the difference between $g_{13}$ and $g_{23}$, and the difference between $\gamma_1$ and $\gamma_2$ are both less than 5 percent of their average values. Therefore, we can safely neglect the difference between dissipation rates $\gamma_1$ and $\gamma_2$, and the difference between coupling rates $g_{13}$ and $g_{23}$.

\begin{figure*}[hbt]
\includegraphics[width=\linewidth]{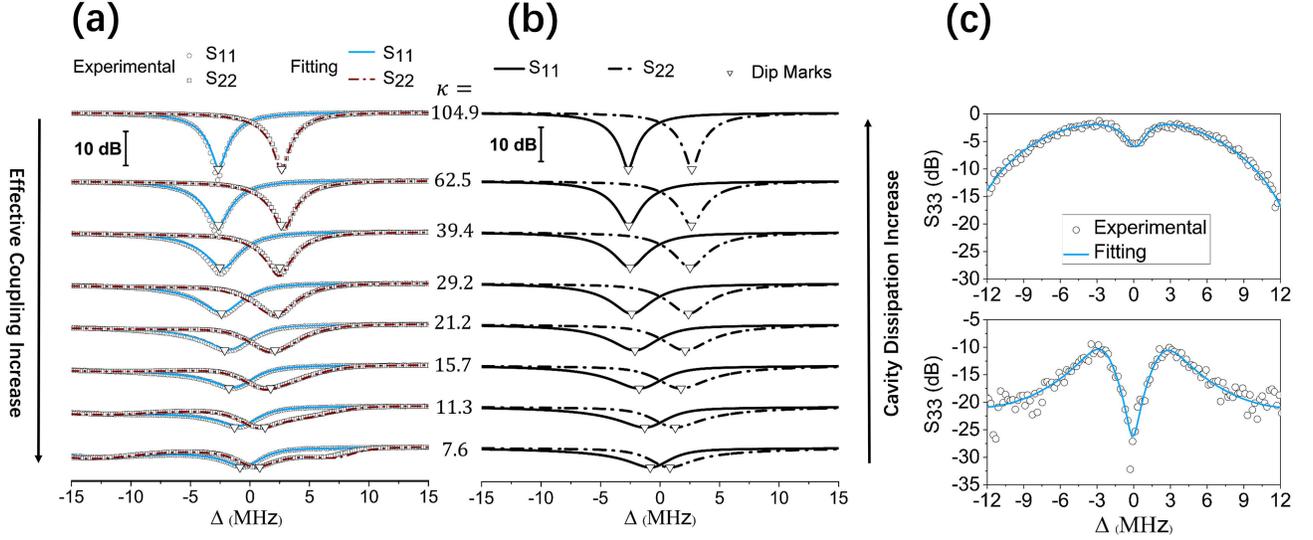}
\caption{\label{fig3}(color online) (a). The magnon-readout results with different cavity dissipation rate $\kappa$ in unit of $\rm MHz$. The circles and the squares present the experiment data of spectrum $\rm S_{11}$ and $\rm S_{22}$, respectively. The solid lines and the dash-dot lines are the fitting results solved by the original experiment Hamiltonian. The triangles mark the resonant dip positions in the original anti-$\mathcal{PT}$ Hamiltonian solved spectra, as shown in (b). (b). The spectra $\rm S_{11}$ and $\rm S_{22}$ solved by the original anti-$\mathcal{PT}$ Hamiltonian in Eq.~(\ref{eff-Hamiltonian}). The triangles indicate the resonant dips in the spectra. (c). Cavity-readout result in anti-$\mathcal{PT}$ symmetry phase (upper panel) and in anti-$\mathcal{PT}$ symmetry broken phase (lower panel). The circles are experimental data and the solid lines are theoretical predictions from the original experiment Hamiltonian with best-fit parameters.}
\end{figure*}

Using the magnon-readout method, we read the reflection parameters $\rm S_{11}$ and $\rm S_{22}$ from antenna 1 and antenna 2, respectively. In this case, the magnon readout antennae coupled to the YIG spheres and to the cavity simultaneously. In other words, the applied probe microwave signal through antenna 1 (2) drives not only the magnon 1 (2) but also the cavity with a relative phase $\varphi_{13}$ ($\varphi_{23}$) simultaneously. The reflected signals from the magnon 1 (2) and the cavity also preserve the same relative phase $\varphi_{13}$ ($\varphi_{23}$). Based on the mechanism, we can solve the input and output field relation as $s_{\rm out} = -s_{\rm in} + \sqrt{\kappa_k}e^{i\varphi_{k3}}c+\sqrt{\gamma_{k1}}a$ ($k=1,\ 2$). Comparing with the magnon-readout method, the cavity-readout method is much simpler. The injected signal from the antenna 3 only drives the cavity, and the input-output field relation preserves the normal form, $s_{\rm out} = -s_{\rm in} + \sqrt{\kappa_3}c$. Using the magnon-cavity-magnon coupled original Hamiltonian and the input-output field relation, we can solve the whole spectra with the standard input-output theory in both readout methods, as shown in Supplementary Material B.

Based on the magnon-readout method, we can demonstrate that the approximation used in our system is valid and construct the anti-$\mathcal{PT}$ symmetry. We apply bias magnetic fields $B_1$ and $B_2$ to bias the magnon 1 and 2 at frequency at $\omega_1$ and $\omega_2$, respectively. In our experiment, the resonant frequencies $\omega_1$ and $\omega_2$ are set to satisfy the relationship $\omega_1-\omega_2=5.4\  \rm MHz$, thus the effective detuning in this configuration is $\left|\Omega\right|=2.7\  \rm MHz$, which is constant in all experiments. And then, we measure the reflection parameters $\rm S_{11}$ and $\rm S_{22}$ from antenna 1 and antenna 2, respectively. As shown in FIG.~\ref{fig3}~(a), the measured $\rm S_{11}$ and $\rm S_{22}$ data are fitted well with the calculated spectra using the original experiment Hamiltonian, as shown in Supplementary Materials B. This result proves that the physical model used in solving the measurement spectra is sufficient. In the other side, the anti-$\mathcal{PT}$ Hamiltonian in Eq.~(\ref{eff-Hamiltonian}) describes a system with dissipatively coupled detuned resonators. We can solve the corresponding reflection spectra with the standard anti-$\mathcal{PT}$ Hamiltonian in Eq.~(\ref{eff-Hamiltonian}), as shown in FIG.~\ref{fig3}~(b), in which the resonant dips are marked with triangles. In order to compare the experimental results with the spectra predicted by the standard anti-$\mathcal{PT}$ Hamiltonian, we draw the triangles at the same position in FIG.~\ref{fig3}~(a). We conclude from this comparison that the resonance occurs at the right frequency and amplitude which are predicted by the standard anti-$\mathcal{PT}$ Hamiltonian. The measurement data demonstrate that the approximations are sufficient and indicate that we successfully construct the anti-$\mathcal{PT}$ symmetry in a magnon-cavity-magnon coupled system.

Based on the cavity-readout method, we can only probe the system through the antenna 3. As shown in FIG.~\ref{fig3}~(c), although the measured data can be fitted well with the spectra given by the original experiment Hamiltonian, the results cannot prove that we successfully construct an anti-$\mathcal{PT}$ system. Because the cavity mode $c$ is eliminated in the large dissipation rate approximation, we cannot compare the measurement results with those obtained by the anti-$\mathcal{PT}$ Hamiltonian.

We now discuss the spontaneous phase transition of the anti-$\mathcal{PT}$ system. In our experiments, the coupling rates between magnons and the cavity are fixed values, which are around 6.5 MHz. In order to observe the anti-$\mathcal{PT}$ symmetry phase transition, we need to increase the effective coupling rate $\Gamma = g^2/\kappa$ by decreasing the overall dissipation rate of the cavity $\kappa$, where $\kappa=\kappa_{\rm int}+\kappa_1+\kappa_2+\kappa_3$ in the magnon-readout, $\kappa_{\rm int}$ is the intrinsic dissipation rate of the cavity (without additional ports), $\kappa_1$, $\kappa_2$ and $\kappa_3$ are the dissipation rates introduced by the antenna 1, antenna 2 and antenna 3, respectively. With different cavity dissipation rates, we obtain the corresponding transmission spectra $\rm S_{11}$ and $\rm S_{22}$, as shown in FIG.~\ref{fig3}~(a). When the cavity dissipation rate $\kappa$ is large, the corresponding effective coupling rate is smaller than the effective magnon detuning (i.e., $\Gamma< \Omega$). The system works in the anti-$\mathcal{PT}$ symmetry broken phase, and the separation between two dips in the spectrum is larger than the full width at half maximum (FWHM). Using the definition of EP, we can obtain the corresponding cavity dissipation rate $\kappa_0 = 15.8\ \rm MHz$. Continuously decreasing the cavity dissipation across the EP results in two main counterintuitive phenomena: (i) decreasing the cavity loss, the measured spectra show mode attraction; (ii) increasing the effective coupling strength between the magnon modes, we observe the energy attraction instead of the mode splitting. These two counterintuitive phenomena are basically induced by the broken anti-$\mathcal{PT}$ symmetry phase transition. When our system works in anti-$\mathcal{PT}$ symmetry phase, the separation between two dips is smaller than the full width at half maximum (FWHM). In order to formulate the relationship between the dip separation and the FWHM, we can define the combined spectrum by $\rm \bar{S}=(S_{11}+S_{22})/2$, as shown in Supplementary Material C.

\begin{figure}[t]
\includegraphics[width=\linewidth]{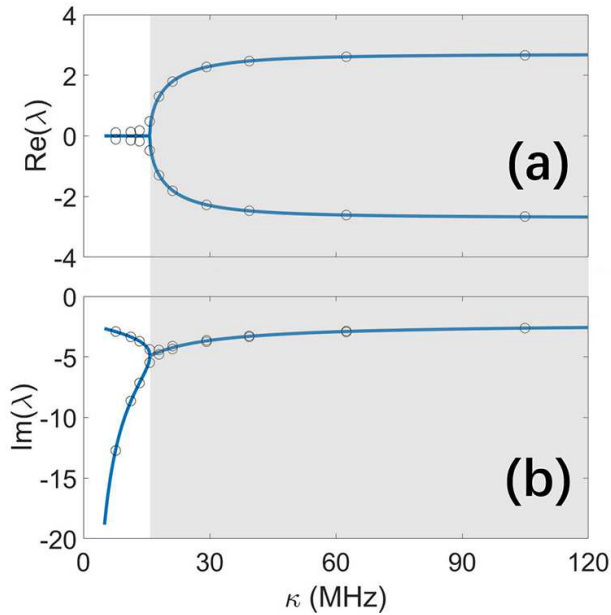}
\caption{\label{fig4}The real part (panel a) and imaginary part (panel b) of the eigenvalues as a function of $\kappa$. The shadow area with $\kappa>15.8\ \rm MHz$ indicates the parametric regime of anti-$\mathcal{PT}$ symmetry broken phase. The experimental data are extracted from the data in TABLE. \ref{sys param} using the method presented in Supplementary Note D.}
\end{figure}

The anti-$\mathcal{PT}$ symmetry induced level attraction can be expressed even more clearly by examining the eigenvalues of different dissipation rate $\kappa$. Using the method elaborated in Supplementary Material D, we extract the eigenvalues and plot the real and imaginary parts as a function of $\kappa$ in FIG.~\ref{fig4}~(a) and (b) respectively, which show excellent agreement with theoretical results. The experimental data in FIG.~\ref{fig4}~(a) reveal that the exceptional point occurs at $\kappa_0 = 15.8\ \rm MHz$, which corresponds to a dissipative coupling rate $\Gamma=2.7\ \rm MHz$. According to Eq.~(\ref{eff-Hamiltonian}), the two real parts of eigenvalues should be $\pm 2.7\ \rm MHz$ when the cavity dissipation rate $\kappa$ approaches infinity. As shown in FIG.~\ref{fig4}~(a), the real parts of eigenvalues corresponding to $\kappa=105\ \rm MHz$, are approximate $\pm 2.7\ \rm MHz$, which are compatible with the theoretical results. When we decrease the value of $\kappa$, the difference between two real parts becomes smaller and is reduced to zero at the EP. The theory predicts that there should be two different imaginary parts in anti-$\mathcal{PT}$ symmetry regime, and a single value of imaginary part in symmetry broken regime. We have also observed this phenomenon in our experiment, as shown in FIG.~\ref{fig4}~(b).

In conclusion, we have successfully constructed anti-$\mathcal{PT}$ symmetry in a magnon-cavity-magnon coupled system without any gain medium, and observed anti-$\mathcal{PT}$ symmetry from the magnon side. From the magnon-readout results, we have observed the broken anti-$\mathcal{PT}$ symmetry at the phase transition point (i.e., the EP), resulting in a counterintuitive energy attraction phenomenon instead of the energy repulsion widely reported in strongly coupled-resonator systems \cite{goryachev2014high, zhang2014strongly, tabuchi2014hybridizing, zhang2015cavity}. Encircling around such exceptional point in the future may allow us to observe various topological operations based on non-adiabatic transitions. The negative frequencies (negative-energy modes) in anti-$\mathcal{PT}$ symmetric Hamiltonian equivalent to harmonic oscillators with negative mass also have a close connection to evade quantum measurement backaction \cite{bernier2018level}. Comparing with the cavity-readout results, we uncover the unique ability of magnon-readout method in exploring multi-magnon-cavity couped systems. Our experiment illustrates the power of the magnon-readout, and motivates further explorations on macroscopic quantum phenomena and the fundamental limit on the quantum sensing based on EPs \cite{wiersig2014enhancing}.

This work was supported by the National Key R\&D Program of China (Grant No. 2018YFA0306600), the CAS (Grants No. GJJSTD20170001 and No. QYZDY-SSW-SLH004), and Anhui Initiative in Quantum Information Technologies (Grant No. AHY050000).

\bibliography{anti-PT}

\end{document}



\title{Supplementary Materials for Observation of anti-$\mathcal{PT}$ symmetry phase transition in the magnon-cavity-magnon coupled system}

\author{Jie Zhao}
\altaffiliation{These authors contributed equally to this work}
\affiliation{Hefei National Laboratory for Physical Sciences at the Microscale and Department of Modern Physics, University of Science and Technology of China, Hefei 230026, China
}
\affiliation{CAS Key Laboratory of Microscale Magnetic Resonance, University of Science and Technology of China, Hefei 230026, China}
\affiliation{Synergetic Innovation Center of Quantum Information and Quantum Physics, University of Science and Technology of China, Hefei 230026, China}

\author{Yulong Liu}
\altaffiliation{These authors contributed equally to this work}
\affiliation{Department of Applied Physics, Aalto University, P.O. Box 15100, FI-00076 Aalto, Finland}

\author{Longhao Wu}
\affiliation{Hefei National Laboratory for Physical Sciences at the Microscale and Department of Modern Physics, University of Science and Technology of China, Hefei 230026, China
}
\affiliation{CAS Key Laboratory of Microscale Magnetic Resonance, University of Science and Technology of China, Hefei 230026, China}
\affiliation{Synergetic Innovation Center of Quantum Information and Quantum Physics, University of Science and Technology of China, Hefei 230026, China}

\author{Chang-Kui Duan}
\affiliation{Hefei National Laboratory for Physical Sciences at the Microscale and Department of Modern Physics, University of Science and Technology of China, Hefei 230026, China
}
\affiliation{CAS Key Laboratory of Microscale Magnetic Resonance, University of Science and Technology of China, Hefei 230026, China}
\affiliation{Synergetic Innovation Center of Quantum Information and Quantum Physics, University of Science and Technology of China, Hefei 230026, China}

\author{Yu-xi Liu}
\affiliation{Institute of Microelectronics, Tsinghua University, Beijing 100084, China}

\author{Jiangfeng Du}
\email[]{djf@ustc.edu.cn}
\affiliation{Hefei National Laboratory for Physical Sciences at the Microscale and Department of Modern Physics, University of Science and Technology of China, Hefei 230026, China
}
\affiliation{CAS Key Laboratory of Microscale Magnetic Resonance, University of Science and Technology of China, Hefei 230026, China}
\affiliation{Synergetic Innovation Center of Quantum Information and Quantum Physics, University of Science and Technology of China, Hefei 230026, China}


\date{\today}

%

\pacs{xxxxxx}

\maketitle


\section{A. Effective anti-$\mathcal{PT}$ Hamiltonian of the magnon-cavity-magnon coupled system}

In our system, two detuning magnons are coupled to a microwave cavity mode separately, and there is no any direct interaction between these two magnons. In this section, we derive the effective Hamiltonian which, describes the effective coupling between two magnons. Based on the effective Hamiltonian, we obtain the anti-$\mathcal{PT}$ Hamiltonian.

In our system, the two YIG spheres are working at low excitation regime and the collective spin excitation (magnon) of YIG spheres can be simply regarded as harmonic oscillators. The original experiment Hamiltonian of the system can be given as

\begin{equation}
H=\omega_1 a^{\dagger}a + \omega_2 b^{\dagger}b + \omega_3 c^{\dagger}c + g_{13}(ac^{\dagger}+a^{\dagger}c) + g_{23}(bc^{\dagger}+b^{\dagger}c).
\end{equation}

\noindent Where we have assumed that $\hbar=1$. $c$ $\left( c^{\dag }\right) $ is the annihilation (creation) operator of the cavity field with resonance frequency $\omega _{3}$. $a$ $\left( a^{\dag }\right) $ is the annihilation (creation) operator of the first magnon mode, and $b$ $\left( b^{\dag}\right) $ is the annihilation (creation) operator of the second magnon mode. $\omega _{1}$ and $\omega _{2}$ are the corresponding resonance frequencies of these two magnon modes. $g_{13}$ and $g_{23}$ represent the single-photon coupling strength between the cavity and the magnon modes.

The corresponding semiclassical Langevin equations are given by

\begin{eqnarray}
\dot{a} &=& -(i\omega_1 + \gamma_1)a-ig_{13}c, \label{S2} \\
\dot{b} &=& -(i\omega_2 + \gamma_2)b-ig_{23}c, \label{S3} \\
\dot{c} &=& -(i\omega_3 + \kappa)c-ig_{13}a-ig_{23}b. \label{S4}
\end{eqnarray}

Introducing the slowly varying amplitudes $A$, $B$ and $C$ with

\begin{eqnarray}
a &=& Ae^{-i\omega_1t}, \\
b &=& Be^{-i\omega_2t}, \\
c &=& Ce^{-i\omega_3t},
\end{eqnarray}

\noindent we use Eqs. (\ref{S2}), (\ref{S3}) and (\ref{S4}) to derive the equations of motion for slowly varying amplitudes as

\begin{eqnarray}
\dot{A} &=& -\gamma_1 A-ig_{13}Ce^{-i\Delta_{13}t}, \label{S8} \\
\dot{B} &=& -\gamma_2 B-ig_{23}Ce^{-i\Delta_{23}t}, \label{S9} \\
\dot{C} &=& -\kappa C-ig_{13}Ae^{i\Delta_{13}t}-ig_{23}Be^{i\Delta_{23}t},
\end{eqnarray}

\noindent where $\Delta_{13} = \omega_3-\omega_1$ and $\Delta_{23}=\omega_3-\omega_2$, which are the frequency detunings between the cavity and the first or the second magnon mode. We can obtain the formal solution of $C$ as

\begin{equation}
C(t)=-ig_{13} \int_{0}^{t} dt^{\prime} A(t^{\prime})e^{i\Delta_{13}t^{\prime}}e^{-\kappa(t-t^{\prime})}-ig_{23}\int_0^t dt^{\prime}B(t^{\prime})e^{i\Delta_{23}t^{\prime}}e^{-\kappa(t-t^{\prime})}
\end{equation}

If the dissipation rate of the cavity mode $c$ is large enough and satisfies the condition $\kappa \gg \gamma_1, \gamma_2$, the amplitude changes of mode $a$ and mode $b$ are small within the range of the integration of the cavity mode $c$. In this case, we can set $A(t^{\prime})=A(t)$ and $B(t^{\prime})=B(t)$, and then we take them out of the integral and get

\begin{equation}
C(t)=\frac{-ig_{13}}{\kappa-i\Delta_{13}}A(t)e^{i\Delta_{13}t} + \frac{-ig_{23}}{\kappa-i\Delta_{23}}B(t)e^{i\Delta_{23}t}.
\end{equation}

\noindent Substituting this equation into Eq. (\ref{S8}) and Eq. (\ref{S9}), we adiabatically eliminate the variables of the mode $c$. The corresponding equations of motion for the mode $a$ and $b$ are then reduced to

\begin{eqnarray}
\dot{A(t)} &=& -\gamma_1A(t)-\frac{g_{13}^2}{\kappa-i\Delta_{13}}A(t)-\frac{g_{13}g_{23}}{\kappa-i\Delta_{23}}B(t)e^{-i(\omega_2-\omega_1)t}, \\
\dot{B(t)} &=& -\gamma_2B(t)-\frac{g_{23}^2}{\kappa-i\Delta_{23}}B(t)-\frac{g_{13}g_{23}}{\kappa-i\Delta_{13}}A(t)e^{-i(\omega_1-\omega_2)t}
\end{eqnarray}

Combine these equations with $A(t)=a(t)e^{i\omega_1t}$ and $B(t)=b(t)e^{i\omega_2t}$, we can obtain

\begin{equation}
i\frac{d}{dt}\left[
\begin{matrix}
a \\
b
\end{matrix}
\right]=\left[
\begin{matrix}
\omega_1-i\left(\gamma_1+\frac{g_{13}^2}{\kappa-i\Delta_{13}}\right) & -i\frac{g_{13}g_{23}}{\kappa-i\Delta_{23}} \\
-i\frac{g_{13}g_{23}}{\kappa-i\Delta_{13}} & \omega_2-i\left(\gamma_2+\frac{g_{23}^2}{\kappa-i\Delta_{23}}\right)
\end{matrix}
\right]
\left[
\begin{matrix}
a \\
b
\end{matrix}
\right]
\end{equation}

The effective Hamiltonian can be

\begin{equation}\label{Heff}
H_{\rm eff}=\left[
\begin{matrix}
\omega_1-i\left(\gamma_1+\frac{g_{13}^2}{\kappa-i\Delta_{13}}\right) & -i\frac{g_{13}g_{23}}{\kappa-i\Delta_{23}} \\
-i\frac{g_{13}g_{23}}{\kappa-i\Delta_{13}} & \omega_2-i\left(\gamma_2+\frac{g_{23}^2}{\kappa-i\Delta_{23}}\right)
\end{matrix}
\right]
\end{equation}

And if $\kappa \gg \left|\Delta_{13}\right|, \left|\Delta_{23}\right|$, the dissipation rates of two magnons equal to each other, i.e., $\gamma_1=\gamma_2=\gamma$, the coupling rates between magnons and cavity are the same, i.e. $g_{13}=g_{23}=g$, the effective Hamiltonian Eq. (\ref{Heff}) is reduced to

\begin{equation}
H_{\rm eff}=\left[
\begin{matrix}
\omega_1-i\left(\gamma+\frac{g^2}{\kappa}\right) & -i\frac{g^2}{\kappa} \\
-i\frac{g^2}{\kappa} & \omega_2-i\left(\gamma+\frac{g^2}{\kappa}\right)
\end{matrix}
\right].
\end{equation}

\noindent Moving to the frame rotating with frequency $\omega=(\omega_1+\omega_2)/2$ and define the effective coupling rate $\Gamma=\frac{g^2}{\kappa}$, the effective Hamiltonian can be reduced to

\begin{equation}\label{anti-PT H}
H_{\rm eff}=\left[
\begin{matrix}
\Omega-i\left(\gamma+\Gamma\right) & -i\Gamma \\
-i\Gamma & -\Omega-i\left(\gamma+\Gamma\right)
\end{matrix}
\right],
\end{equation}

\noindent where $\Omega = (\omega_1 - \omega_2)/2$ is the effective detuning in the rotating frame. The effective Hamiltonian in Eq. (\ref{anti-PT H}) is anti-$\mathcal{PT}$ symmetric \cite{yang2017anti}.

\section{B. Calculation of the transmission spectra}
In our experiment, the antenna 1 (antenna 2) is coupled to magnon 1 (magnon 2) and is used to readout the transmission spectra of the system from the magnons. The antennae are not only coupled to the magnons but also coupled to the cavity. As discussed in the main text, the antenna 1 - cavity (antenna 2 - cavity) coupling introduces the dissipation rate $\kappa_1$ ($\kappa_2$) to the cavity. When we apply a signal to drive the magnon, this signal also drives the cavity with a relative phase $\varphi_{13}$ ($\varphi_{23}$). When we measure the reflected signal, the signal coming out from the cavity also must be taken into account. Following this idea, the transmission spectra can be solved using the standard input-output theory. In this section, we solve the transmission spectrum $\rm S_{11}$, which is read from the magnon 1. The same method can be applied to the calculation of the spectrum $\rm S_{22}$.

When we measure the transmission spectrum $\rm S_{11}$, a microwave pulse with amplitude $s$ and frequency $\omega_p$ is injected into the antenna 1. This microwave pulse drives the magnon 1 and the cavity simultaneously with a relative phase $\varphi_{13}$. In this situation, the system Hamiltonian can be

\begin{eqnarray}
\nonumber
H &=& \omega_1 a^{\dagger}a + \omega_2 b^{\dagger}b + \omega_3 c^{\dagger}c + g_{13}(ac^{\dagger}+a^{\dagger}c) + g_{23}(bc^{\dagger}+b^{\dagger}c) \\
&+& i\sqrt{\gamma_{11}}s(a^{\dagger}e^{-i\omega_pt}+h.c.)+ i\sqrt{\kappa_1}s(c^{\dagger}e^{-i\omega_pt-i\varphi_{13}}+h.c.),
\end{eqnarray}

\noindent where $\gamma_{11}$ is the antenna induced magnon dissipation rate and $\kappa_1$ is the antenna induced cavity dissipation rate. The coupling terms in our experiments are actually $g_{13}(ac^{\dagger}+ e^{i\Phi_{13}}a^{\dagger}c) + g_{23}(bc^{\dagger}+ e^{i\Phi_{23}}b^{\dagger}c)$ \cite{PhysRevLett.121.137203}. The measured values of $\Phi_{13}$ and $\Phi_{23}$ are around $0.03\pi$, which is much smaller than $\pi$, and the calculated spectra fit well with the experimental data. Therefore, we omit the phase $\Phi_{13}$ and $\Phi_{23}$. In the rotating reference frame with the frequency $\omega_p$, the Hamiltonian is

\begin{eqnarray}
\nonumber
H &=& \Delta_1 a^{\dagger}a + \Delta_2 b^{\dagger}b + \Delta_3 c^{\dagger}c + g_{13}(ac^{\dagger}+a^{\dagger}c) + g_{23}(bc^{\dagger}+b^{\dagger}c) \\
&+& i\sqrt{\gamma_{11}}s(a^{\dagger}+h.c.)+ i\sqrt{\kappa_1}s(c^{\dagger}e^{-i\varphi_{13}}+h.c.),
\end{eqnarray}

\noindent where $\Delta_i=\omega_i-\omega_p, i=1, 2, 3$. The corresponding semiclassical Langevin equations with zero mean value of noise operators are

\begin{eqnarray}
\dot{a} &=& -(i\Delta_1+\gamma_1)a-ig_{13}c+\sqrt{\gamma_{11}}s, \\
\dot{b} &=& -(i\Delta_2+\gamma_2)b-ig_{23}c, \\
\dot{c} &=& -(i\Delta_3+\kappa)c-ig_{13}a-ig_{23}b+\sqrt{\kappa_1}se^{-i\varphi_{13}},
\end{eqnarray}

\noindent where $\kappa$ is the overall dissipation rate of the cavity, $\gamma_1$ and $\gamma_2$ are the overall dissipation rates of the magnon 1 and magnon 2, respectively. We have defined that $o\equiv \left<o\right>$, with $o=a,\ b,\ c$.

\noindent Using the Langevin equations, we can obtain the steady state solution of $a$ and $c$:

\begin{eqnarray}
a &=& \frac{\sqrt{2\gamma_{11}}s}{i\Delta_1+\gamma_1+\frac{g_{13}^2}{i\Delta_3+\kappa+\frac{g_{23}^2}{i\Delta_2+\gamma_2}}} - \frac{\frac{ig_{13}\sqrt{2\kappa_1}se^{-i\varphi_{13}}}{i\Delta_3+\kappa+ \frac{g_{23}^2}{i\Delta_2+\gamma_2}}}{i\Delta_1+\gamma_1+\frac{g_{13}^2}{i\Delta_3+\kappa+\frac{g_{23}^2}{i\Delta_2+\gamma_2}}} \\
c &=& \frac{\sqrt{2\kappa_1}se^{-i\varphi_{13}}-\frac{ig_{13}\sqrt{2\gamma_{11}}s}{i\Delta_1+\gamma_1}}{i\Delta_3+\kappa+ \frac{g_{13}^2}{i\Delta_1+\gamma_1}+\frac{g_{23}^2}{i\Delta_2+\gamma_2}}
\end{eqnarray}

Following the method presented in Refs. \cite{walls2007quantum} and \cite{clerk2010introduction}, we can solve the boundary condition, which describes the relationship between the external fields and the intracavity fields. We first consider the output field $a_{\rm out}$, the boundary condition is,

\begin{equation}\label{a-out}
a_{\rm out} = -s + \sqrt{\gamma_{11}}a+\sqrt{\kappa_1}ce^{i\varphi_{13}}.
\end{equation}

\noindent Similarly, the boundary condition related to the output field $c_{\rm out}$ can be obtained,

\begin{equation}\label{c-out}
c_{\rm out} = -s + \sqrt{\gamma_{11}}a+\sqrt{\kappa_1}ce^{i\varphi_{13}}.
\end{equation}

\noindent The overall output field can be obtained by adding the $a_{\rm out}$ part and the $c_{\rm out}$ part. Because the input field $s$ is added twice, the reflection coefficient can be obtained as

\begin{eqnarray}
\nonumber
t_1 &=& \frac{a_{\rm out}+c_{\rm out}}{2s} \\
\nonumber
&=& -1+\frac{2\gamma_{11}}{i\Delta_1+\gamma_1+\frac{g_{13}^2}{i\Delta_3+\kappa+\frac{g_{23}^2}{i\Delta_2+\gamma_2}}} - \frac{\frac{i2g_{13}\sqrt{\gamma_{11}\kappa_1}e^{-i\varphi_{13}}}{i\Delta_3+\kappa + \frac{g_{23}^2}{i\Delta_2+\gamma_2}}}{i\Delta_1+\gamma_1+\frac{g_{13}^2}{i\Delta_3+\kappa+\frac{g_{23}^2}{i\Delta_2+\gamma_2}}} \\
&+&\frac{2\kappa_1e^{-i\varphi_{13}}-\frac{i2g_{13}\sqrt{\gamma_{11}\kappa_1}}{i\Delta_1+\gamma_1}}{i\Delta_3+\kappa+ \frac{g_{13}^2}{i\Delta_1+\gamma_1}+\frac{g_{23}^2}{i\Delta_2+\gamma_2}}e^{i\varphi_{13}}.
\end{eqnarray}

Using the reflection coefficient $t_1$, we can easily solve the S parameter ${\rm S_{11}}=\left|t_1\right|$. It's easily to verify that the S parameter $\rm S_{22}$ can be solved with the same method, ${\rm S_{22}}=\left|t_2\right|$, where

\begin{eqnarray}
\nonumber
t_2 &=& \frac{b_{\rm out}+c_{\rm out}}{2s} \\
\nonumber
&=& -1+\frac{2\gamma_{21}}{i\Delta_2+\gamma_2+\frac{g_{23}^2}{i\Delta_3+\kappa+\frac{g_{13}^2}{i\Delta_1+\gamma_1}}} - \frac{\frac{i2g_{23}\sqrt{\gamma_{21}\kappa_2}e^{-i\varphi_{23}}}{i\Delta_3+\kappa + \frac{g_{13}^2}{i\Delta_1+\gamma_1}}}{i\Delta_2+\gamma_2+\frac{g_{23}^2}{i\Delta_3+\kappa+\frac{g_{13}^2}{i\Delta_1+\gamma_1}}} \\
&+&\frac{2\kappa_2e^{-i\varphi_{23}}-\frac{i2g_{23}\sqrt{\gamma_{21}\kappa_2}}{i\Delta_2+\gamma_2}}{i\Delta_3+\kappa+ \frac{g_{13}^2}{i\Delta_1+\gamma_1}+\frac{g_{23}^2}{i\Delta_2+\gamma_2}}e^{i\varphi_{23}}.
\end{eqnarray}

Using the obtained equations of $\rm S_{11}$ and $\rm S_{22}$, we can fit the experiment data, as shown in Fig. 2a in the main text. In the fitting process of $\rm S_{11}$ or $\rm S_{22}$, there are only fitting parameters $\varphi_{13}$ or $\varphi_{23}$, respectively.

\section{C. The combined spectra $\rm\bar{S}$}

\begin{figure}
\includegraphics[width=8cm]{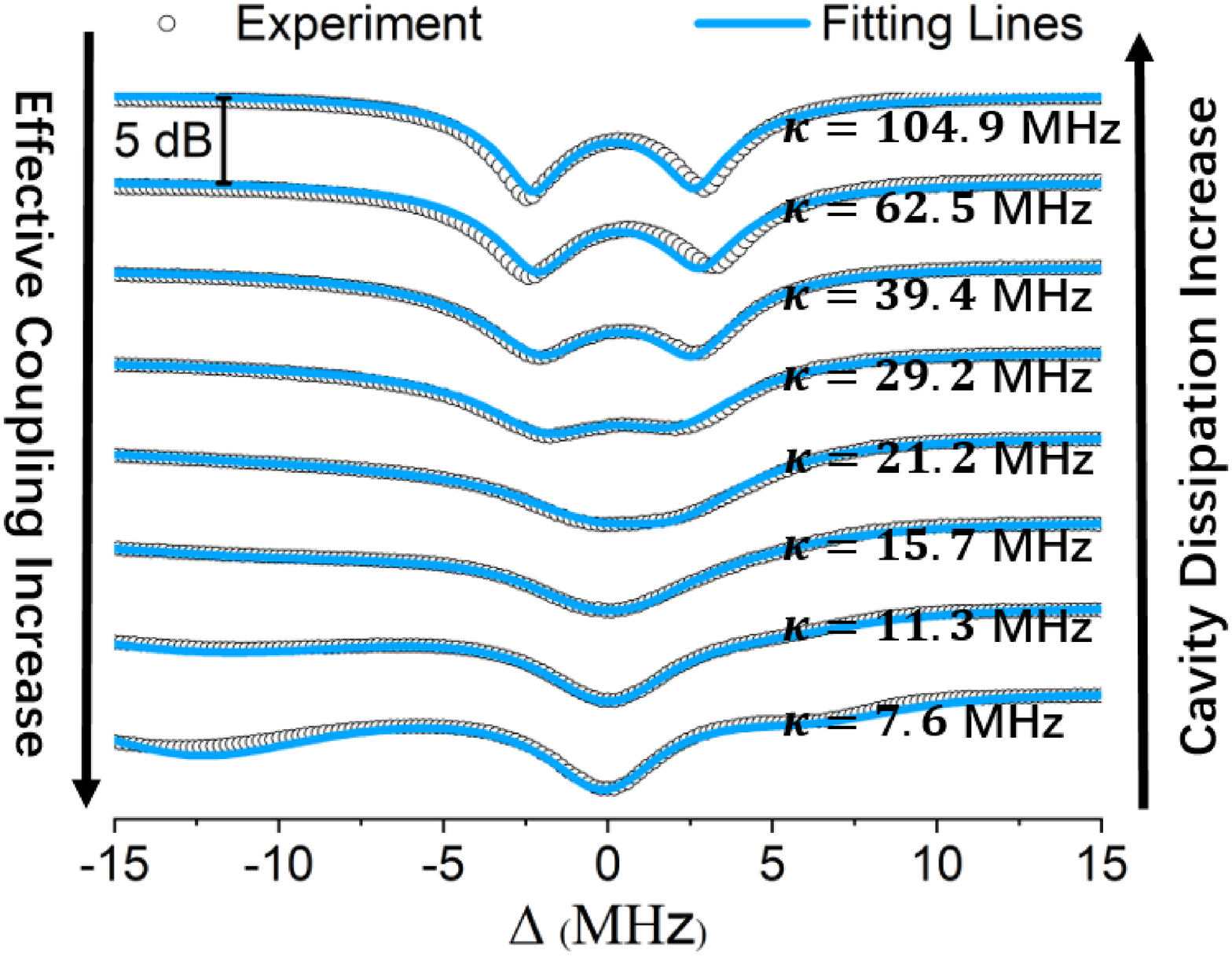}
\caption{\label{figS2}{color online}. The combined spectra with different cavity dissipation rates. }
\end{figure}

In spectroscopy, we think two peaks cannot be distinguished when the separation between them is smaller than the full width at half maximum (FWHM) of each peak. In this situation, we can observe one peak in the spectrum. In our experiment, we can obtain the transmission spectra $\rm S_{11}$ and $\rm S_{22}$ with different separations between resonant dips. In order to conveniently measure the dip separation, we define the combined spectra as $\rm \bar{S} = (S_{11} + S_{22})/2$. In our experiment, the experimentally obtained spectra ${\rm S_{11}}(\omega_p)$, ${\rm S_{22}}(\omega_p)$ and the combined spectrum ${\rm\bar{S}}(\omega_p)$ are fitted very well. Under different cavity dissipation rates, we can obtain the combined spectra as shown in Fig.~\ref{figS2}.

We can find from Fig.~\ref{figS2} that there are two dips in the spectrum when the experiment system works in anti-$\mathcal{PT}$ symmetry broken regime. If the cavity loss is decreased, then the measured spectra also show mode attraction. In Fig.~\ref{figS2}, the anti-$\mathcal{PT}$ symmetry breaking process is illustrated clearer compared with the data shown in Fig.~2~(a) in the main text. However, there is not a physical quantity corresponding to the combined spectrum.

\section{D. The real and imaginary parts of the eigenenergy}

In our experiment, the line shapes of reflection coefficients $\rm S_{11}$ and $\rm S_{22}$ are not the normally Lorentzian ones when the cavity dissipation rate is not large enough. It is not suitable to extract the real part (resonant frequency) or the imaginary part (line width) of the eigenenergy of the system by directly fitting the reflection coefficients. As illustrated in the main text, we experimentally obtain the parameters of the system and theoretically solve the eigenenergies using the following Hamiltonian:

\begin{equation}\nonumber
H_{\rm eff}=\left[
\begin{matrix}
\omega_1-i\left(\gamma_1+\frac{g_{13}^2}{\kappa-i\Delta_{13}}\right) & -i\frac{g_{13}g_{23}}{\kappa-i\Delta_{23}} \\
-i\frac{g_{13}g_{23}}{\kappa-i\Delta_{13}} & \omega_2-i\left(\gamma_2+\frac{g_{23}^2}{\kappa-i\Delta_{23}}\right)
\end{matrix}
\right]
\end{equation}

The fitting lines in Fig. 4 are drawn with the mean value of the corresponding parameters with the anti-$\mathcal{PT}$ Hamiltonian Eq. \ref{anti-PT H}.

\section{E. Data obtained from cavity side}

In most experiments about magnon - cavity polariton, the system is probed from the cavity. In this section, we present the data obtained from the cavity.

The experimental setup is the same as the one used in the main text except that the antenna 1 and antenna 2 are removed. In this setup, the coupling rate between the cavity and magnon 1 (magnon 2) is $g_{13}=9.77\ \rm MHz$ and $g_{23}=9.61\ \rm MHz$. The intrinsic dissipation rates of two magnons and the effective magnon detunings are the same as the values used in the main text. We probe the status of the system by measuring the reflection coefficient $\rm S_{33}$ from antenna 3. In order to maintain the consistency of experimental data obtained with different cavity dissipation rates, we need to keep the antenna 3 critically coupled to the cavity. In our experiment, we change the intrinsic dissipation rate of the cavity by filling it with dissipative materials, and the antenna 3 induced dissipation rate by tuning the length of antenna 3. We tune the intrinsic cavity dissipation rate and the antenna 3 induced dissipation rate at the same time, so that the antenna 3 is critically coupled to the cavity (the reflection coefficient $\rm S_{33}$ at the resonant frequency is less than -20 dB).

\begin{figure}
\includegraphics[width=16cm]{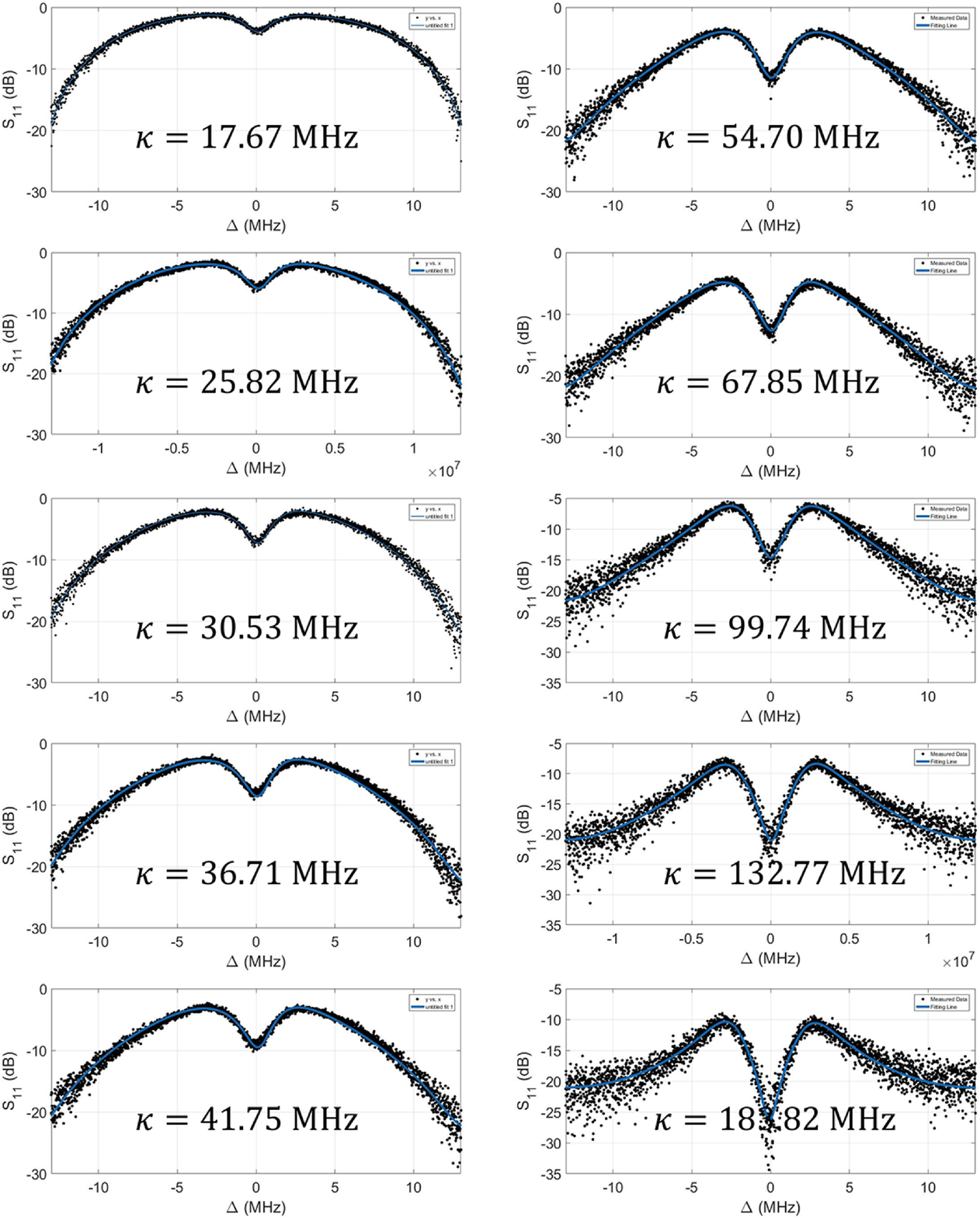}
\caption{\label{figS1}{color online}. The reflection coefficient spectra $\rm S_{33}$ read from the antenna 3 with different cavity dissipation rates. The black dot are experiment data and the blue solid lines are the theorectical fitting data. In this setup, the exceptional point is defined by cavity dissipation rate of around $\kappa_0=34.8\ \rm MHz$. The two peaks tend to merge into a single one when we reduce the value of cavity dissipation rate. However, there are always two peaks in this figure even if the cavity dissipation rate is less than the value of $\kappa_0$.}
\end{figure}

The experiment data are presented in Fig.~\ref{figS1}. In our setup, the exceptional point is defined by the cavity dissipation rate around $\kappa_0=34.8\ \rm MHz$. As illustrated in the main text, there is only one resonant frequency when our system works in the anti-$\mathcal{PT}$ symmetric phase regime (i. e. the cavity dissipation rate is less than the critical value $\kappa_0$). We find in Fig.~\ref{figS1} that although the two peaks tend to merge into a single one when we reduce the value of cavity dissipation rate, the two peaks are separated even if the cavity dissipation rate is less than the value of $\kappa_0$. The experimental data obtained from the cavity cannot reveal the exceptional point schematically. Thus we conclude that the data obtained from the cavity cannot be used as a demonstration of the phase transition of an anti-$\mathcal{PT}$ system.

\bibliography{anti-PT-supp}